\documentclass{moriond}

\def\be{\begin{equation}}
\def\ee{\end{equation}}
\def\bea{\begin{eqnarray}}
\def\eea{\end{eqnarray}}

\usepackage{SIunits}
\usepackage{amsmath}
\newcommand{\Photo}{}
\begin{document}
\vspace*{4cm}
\title{High-accuracy polarimetry for CMB: new frontiers with the POLOCALC project}

\author{A. Novelli, F. Astori, L. Bizzarri, F. Cacciotti, G. Cattaneo, G. Coppi, N. Dachlythra, I. Karaaslan, N. Mezzanzanica, F. Nati, M. Zannoni}

\address{Department of Physics G. Occhialini, University of Milan Bicocca, Piazza della Scienza 3, 20126\\
Milan, Italy}

\maketitle
\abstracts{
Modern telescopes observing the Cosmic Microwave Background (CMB) polarization require an exquisite control of systematics to target Inflationary Gravitational Waves (IGW), Cosmic Birefringence (CB), and Primordial Magnetic Fields (PMF). The absolute polarization angle of the detectors is a critical parameter to disentangle the $E$-modes and $B$-modes of the CMB, allowing a correct detection of primordial $B$-modes as well as testing Cosmic Birefringence theories. 
To this end, we discuss the current status of the POLOCALC project, an ERC Advanced Grant that aims to develop air-borne calibration sources for CMB small-aperture telescopes. The main scientific objective of POLOCALC is to enable a direct calibration of the absolute polarization angle of CMB polarimeters with an accuracy of $0.01 \degree $. We present the latest developments regarding the calibration source, the calibration strategies designed to use drone-based calibrators, and the application to modern ground-based experiments.}

\section{Introduction}
Several current and prospective Cosmic Microwave Background (CMB) experiments are targeting the $B$-modes of the CMB. In order to achieve this goal they need a precise control of the instrument systematics and, in particular, of their absolute polarization angle $\theta_\text{pol}$. A polarization angle miscalibration causes $E$ to $B$ leakage and limits the ability to detect Inflationary Gravitational Waves (IGW) and probe Cosmic Birefringence (CB) theories. Modern experiments such as Simons Observatory or LiteBIRD require arcminute accuracy on $\theta_\text{pol}$ calibration in order to be able to make a detection of the tensor-to-scalar ratio at the level of $r\simeq0.01$ \cite{PhysRevD.77.083003} \cite{MacTavish_2008} \cite{O_Dea_2007}.
This requirement is relaxed if one uses $EB$-nulling techniques to determine the polarization angle of the detectors \cite{Abitbol_2021}. In this case an absolute calibration to the level of $0.2\degree$ is related to an uncertainty on $r$ of the order of $2\times10^{-4}$.
The $EB$-nulling approach, however, works only under the assumption that there are no parity violating mechanisms such as CB, for which there is some preliminary evidence \cite{Minami_2020}.\\
In this paper we discuss the status of POLOCALC \cite{Nati_2017}: a novel way to perform absolute polarization angle calibration using drone-based artificial sources. The main scientific objective of POLOCALC is to enable a direct calibration of $\theta_\text{pol}$ with an accuracy between $0.1\degree$ and $0.01\degree$. In the next sections we will discuss the latest updates in terms of source development, flight-plans and observation strategy, as well as simulations and preliminary analysis.
\section{Calibration source}
There are several requirements that come into play when building a drone-based calibration source for CMB telescopes. One of the most critical is to have a payload which is sufficiently small and light to fly under a drone at high altitude. For this reason all the components of the source are housed inside a custom 3-D printed frame that is smaller than $16\times16\times13\,\centi\meter^3$ and the overall payload weight is kept below $2.5\,\kilo\gram$. This approach is different from the aluminum payload that was used for the first prototype \cite{Coppi_2025}. Site tests have demonstrated that the 3-D printed payload, in addition to being lighter and easier to fabricate, can also guarantee the structural rigidity and passive cooling which were the main drivers in opting for an aluminum frame.\\
Since we aim to calibrate CMB telescopes it is necessary to have a source capable to operate at the typical frequencies used by those telescopes. For this reason, together with the HOVERCAL \cite{Hovercal} collaboration, we developed three versions of the calibration source capable to operate in the frequency bands: $75-115\,\giga\hertz$, $110-170\,\giga\hertz$, and $220-300\,\giga\hertz$.\\
In order to direct the emission of the source towards the telescope we have a set of antennas that can be used, ranging from a more directive $30\,\degree$ antenna to a Near-Field-Probe with a $120\,\degree$ emission. The trade-off consists in the fact that a more directive antenna ensures less diffraction at the wire-grid aperture and a cleaner beam. Having a wider emission, instead, allows one to calibrate at the same time multiple close-by telescopes with comparable power emitted from the calibration source.\\
The last optical element of the calibration source is an $8\,\centi\meter$ wire-grid placed in front of the emitting antenna. The wire-grid has two functions: it cleans the polarization emission of the antenna and it is used to calibrate the source emission with respect to the reconstructed attitude of the payload (see section \ref{sec: attitude reconstruction}).\\
In order to reconstruct the attitude of the payload we have an inclinometer and a ground-facing camera. The inclinometer provides high-rate relative orientation data but it's absolute calibration can be problematic. For this reason, the photogrammetry camera is used to perform photogrammetry of geo-referenced targets and provides the absolute attitude of the payload, even though it does so with a lower sampling rate.
%Finally, in order to make the calibration source airborne and point it towards the telescope we use a Matrix-600 PRO drone from DJI and a Ronin-MX stabilization gimbal.
\section{Attitude Reconstruction}
\label{sec: attitude reconstruction}
In order to use the POLOCALC source as an absolute polarization angle calibrator for CMB telescopes we must be able to reconstruct the attitude of the source during flight. This effort can be divided into two parts. The first part consist of monitoring the position of the source and can be accomplished with $\centi \meter$-level accuracy thanks to the use of a differential GPS.\\
The second part of the attitude reconstruction is more challenging and consists of retrieving the pointing of the source. We achieve that by placing and geo-tagging 40 photogrammetry targets in an area $250 \times250\,\meter$ around the telescope. The targets are $1.5\,\meter$-diameter blue disks that are clearly visible using the camera on-board the payload. Using the target's position in the image and in the 3-D world we can perform photogrammetry and estimate the pointing of the camera with sub-degree accuracy.\\
However, solving for the pointing of the camera does not automatically determine the roll-angle of the wire-grid and the polarization angle of the source. Indeed, the camera and the wire-grid can be mounted with a roll-angle offset with respect to each other. To calibrate this offset before and after each flight we shine a laser from inside the payload through the wire-grid. The laser, interacting with the wires, produces a diffraction pattern that is perpendicular to the wires. By projecting this pattern onto a white wall and photographing it with the photogrammetry camera we can determine the relative roll angle between the wires of the wire-grid and the camera sensor. Applying this correction to the result of photogrammetry we can determine the polarization angle of the source over time.
\section{Deployment}
We deployed and tested the POLOCALC source on Cerro Toco, at the site of the CLASS \cite{Essinger_Hileman_2014} and Simons Observatory (SO) \cite{Ade_2019} telescopes. The calibration strategy is based on flying the drone along a vertical arc centered on the telescope, while continuously pointing the source toward it using the onboard gimbal system. This geometry ensures a nearly constant distance between the source and the telescope, and therefore a stable received power throughout the flight. An arc radius of approximately $500\, \meter$ was found to provide a good compromise between drone operational limits and the need to avoid detector saturation from the drone’s thermal emission.
During operations, the telescope scans in azimuth under nominal observing conditions, while the drone moves vertically along the arc. The combination of these motions allows the calibration source to illuminate the entire focal plane efficiently (Figure \ref{fig: only figure} bottom-right).
\section{Preliminary data}
We can model the passage of the drone in front of a CMB telescope as the sum of the source emission, the drone body emission and the background sky emission.
\begin{align*}
d(t) = & d_\text{sky} + d_\text{drone} + d_\text{source}=\\
=& \frac{1}{2}\left[ I_\text{sky} +P_\text{sky}(1+\cos(4\theta_\text{HWP}+2\theta_\text{sky}))\right]+\\
&+\frac{1}{2}B(t)\left[ I_\text{drone} +P_\text{drone}(1+\cos(4\theta_\text{HWP}+2\theta_\text{drone}))\right]+\\
&+\frac{1}{2}B(t) \,C_\text{chop} \left[ I_\text{source} +P_\text{source}(1+\cos(4\theta_\text{HWP}+2\theta_\text{source}))\right]
\end{align*}
The top term of the three represents the signal that the telescope receives from the sky, characterized by its unpolarized intensity $I$, polarized intensity $P$, and polarization angle $\theta$. The signal is modulated rotating the angle of the HWP of the telescope $\theta_\text{HWP}$. 
Similarly, the second term describes the thermal emission of the drone body. It is multiplied by a beam-profile function $B(t)$ that describes the efficiency with which the drone is observed as it crosses the detector beam.
Finally, the last term describes the source emission as seen by the telescope. This signal is not only affected by the positioning of the drone inside of the beam of the telescope but also by the chopping of the calibration source $C_\text{chop}(t)$.
\begin{figure}
\begin{minipage}{0.5\linewidth}
\centerline{\includegraphics[width=\linewidth]{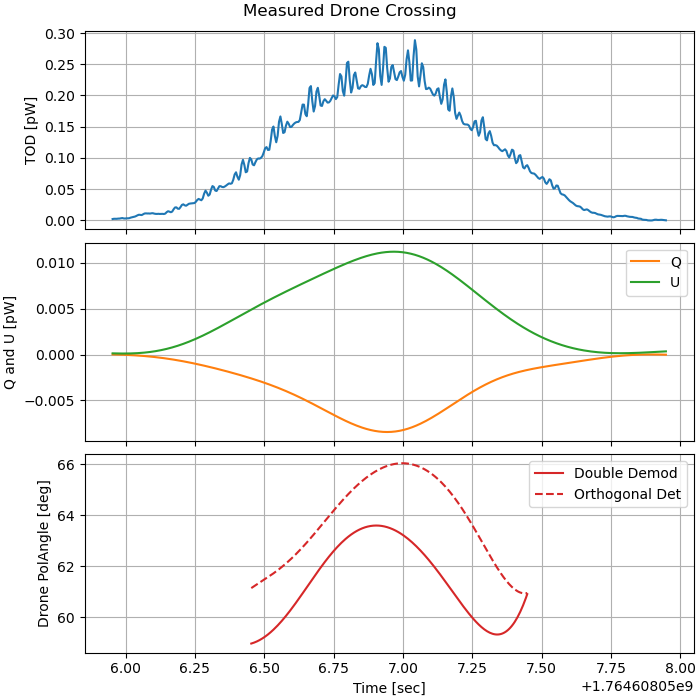}}
\end{minipage}
\hfill
\begin{minipage}{0.50\linewidth}
\centerline{\includegraphics[width=0.7\linewidth]{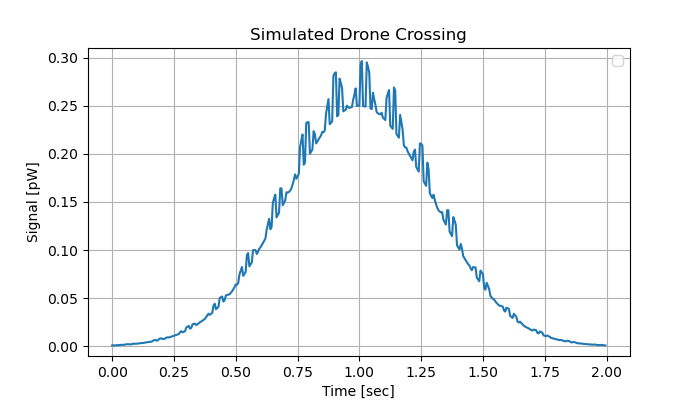}}
\vspace{.3cm}
\centerline{\includegraphics[width=0.8\linewidth]{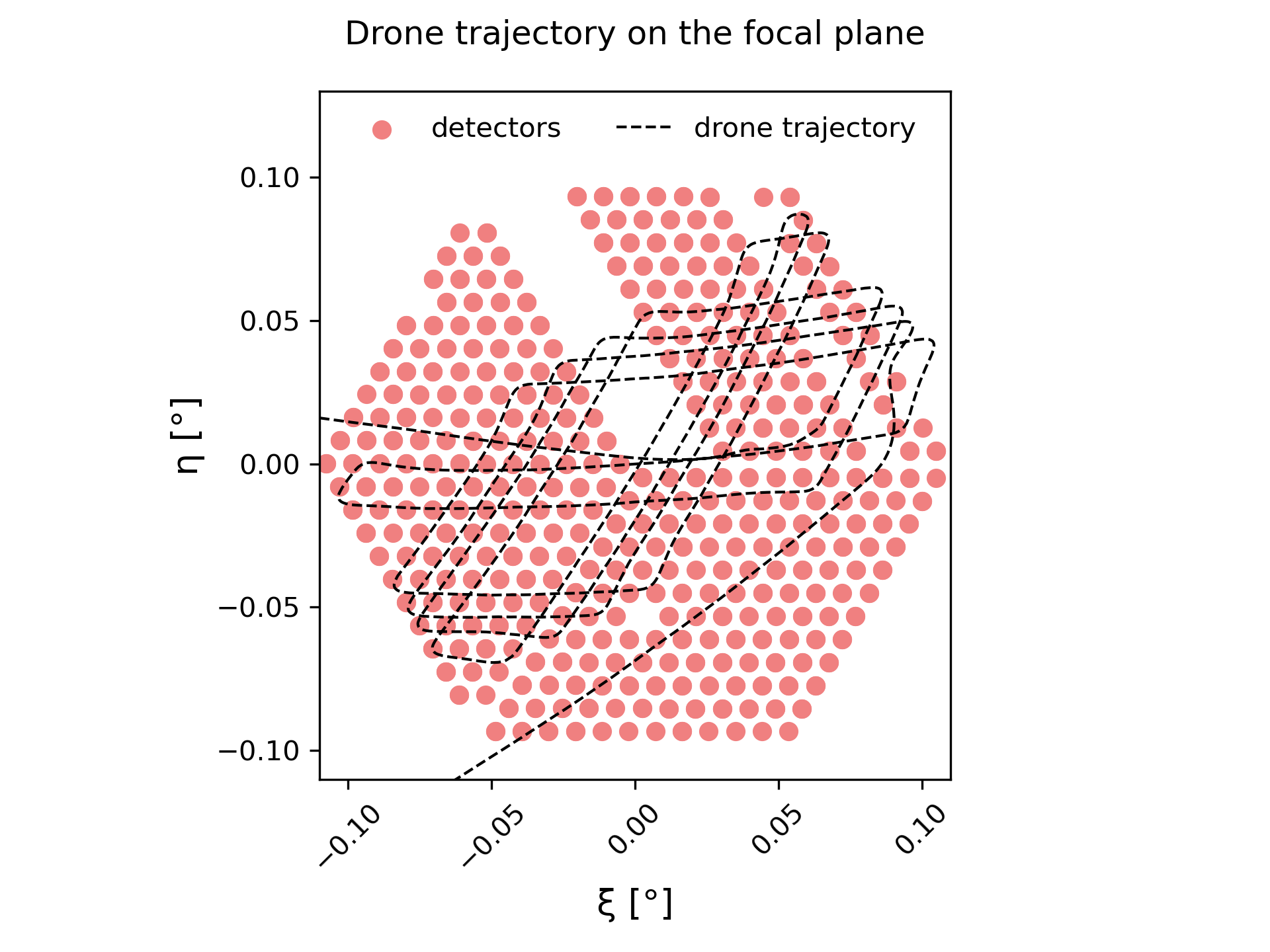}}
\end{minipage}
\caption[]{Left: Measured drone crossing. Top-Right: Simulated drone crossing. Bottom-Right: focal plane coverage in one flight.}
\label{fig: only figure}
\end{figure}\\
In Figure~\ref{fig: only figure} we show the time-ordered data acquired as the drone passes through the detector field of view. The top-right panel presents a simulated timeline generated using the data model described above, while the left panel shows the corresponding measured data. The excellent agreement between simulation and observation provides strong validation of the model.\\
Starting from the raw data, we can implement a double-demodulation pipeline that first isolates the polarized emission of the source and then reconstructs its Stokes $Q$ and $U$ parameters. This allows us to estimate the polarization angle as measured by the detector and compare it to the absolute calibration obtained through photogrammetry.

\section{Future plans}
In the first two years of the project, we have consolidated the payload design and successfully validated flight operations at altitudes up to $5600\,\meter$. We have observed the calibration source with both the Simons Observatory SATs and CLASS telescopes and demonstrated the ability to recover its polarization using the double-demodulation technique. Ongoing analyses include detailed studies of detector beams, chromatic effects, detector efficiency, and absolute polarization angle calibration.\\
To complete the project and reach the target absolute polarization accuracy of $0.1^\circ\div0.01^\circ$, further improvements are required. In particular, we aim to refine the characterization of the source beam and enhance the calibration of the photogrammetry system. This two improvements will allow us to match the photogrammetry information with the telescope data and achieve absolute calibration.
\section*{Acknowledgments}
Novelli and Nati acknowledge funding from the European Union (ERC, POLOCALC, 101096035).
\section*{References}
\bibliography{moriond}

\end{document}